# Femtosecond photo-switching of interface polaritons in black phosphorus heterostructures

*Markus A. Huber[1], Fabian Mooshammer[1], Markus Plankl[1], Leonardo Viti[2], Fabian Sandner[1], Lukas Z. Kastner[1], Tobias Frank[1], Jaroslav Fabian[1], Miriam S. Vitiello[2\*], Tyler L. Cocker[1\*] and Rupert Huber[1]*

[1] Department of Physics, University of Regensburg, 93040 Regensburg, Germany

[2] NEST, CNR – Istituto Nanoscienze and Scuola Normale Superiore, 56127 Pisa, Italy

**The possibility of hybridizing collective electronic motion with mid-infrared (mid-IR) light to form surface polaritons has made van der Waals layered materials a versatile platform for extreme light confinement[1-5] and tailored nanophotonics[6-8]. Graphene[9,10] and its heterostructures[11-14] have attracted particular attention because the absence of an energy gap allows for plasmon polaritons to be continuously tuned. Here, we introduce black phosphorus[15-19] (BP) as a promising new material in surface polaritonics that features key advantages for ultrafast switching. Unlike graphene, BP is a van der Waals bonded semiconductor, which enables high-contrast interband excitation of electron-hole pairs by ultrashort near-infrared (near-IR) pulses. We design a $SiO_2$/BP/$SiO_2$ heterostructure in which the surface phonon modes of the $SiO_2$ layers hybridize with surface plasmon modes in BP that can be activated by photo-induced interband excitation. Within the Reststrahlen band of $SiO_2$, the hybrid interface polariton assumes surface-phonon-like properties, with a well-defined frequency and momentum and excellent coherence. During the lifetime of the photogenerated electron-hole plasma, coherent polariton waves can be launched by a broadband mid-IR pulse coupled to the tip of a scattering-type scanning near-field optical microscopy (s-SNOM) setup. The scattered radiation allows us to trace the new hybrid mode in time, energy, and space. We find that the surface mode can be activated within ~50 fs and disappears within 5 ps, as the electron-hole pairs in BP recombine. The excellent switching contrast and switching speed, the coherence properties, and the constant wavelength of this transient mode make it a promising candidate for ultrafast nanophotonic devices.**

*e-mail: miriam.vitiello@sns.it; tyler.cocker@physik.uni-regensburg.de

The BP flakes studied here were transferred onto a Si/SiO$_2$ substrate (oxide thickness: 300 nm) by mechanical exfoliation[20]. Since BP rapidly degrades under ambient conditions[21] it must be encapsulated under a protective layer to prevent oxidation[22,23]. We deposited a 5 nm protective layer of SiO$_2$ over the entire sample by sputtering. The protected BP flakes remained stable and clean for months. A schematic of a SiO$_2$/BP/SiO$_2$ heterostructure is shown in Fig. 1a, alongside a typical optical microscope image. We explore the system with ultrafast near-IR pump / mid-IR probe s-SNOM, which combines the strengths of near-field microscopy with femtosecond time resolution and has successfully accessed the ultrafast local photoconductivity of graphene[9], semiconductors[24], and insulator-metal phase change materials[25], as well as transient surface plasmon modes on graphene[10].

Our SiO$_2$/BP/SiO$_2$ heterostructure is globally illuminated with a near-IR pump pulse (centre wavelength 1560 nm, pulse duration 40 fs FWHM), which drives interband excitations in the BP flake (Fig. 1b) but does not affect the large-bandgap SiO$_2$ layers. The photo-generated free carrier density defines a plasma frequency in the mid-IR. As a consequence, the scattered near-field intensity $I_4$ (for further details on the demodulation technique see Methods) of our mid-IR probe pulses is strongly modified when the near-field tip is located over the BP heterostructure (Fig. 1c). The signal sets on rapidly, for example with a rise time of ~380 fs (from 20% to 80% of its maximum) for the heterostructure shown in Fig. 1, followed by a non-exponential decay for pump-probe delay times $t_{pp}$ < 5 ps and a persistent state of reduced scattering response that survives for >10 ps. These dynamics are related to the ultrafast evolution of the BP complex conductivity – and hence the dynamics of electron-hole pair generation and recombination – but the nonlinear scattering response makes this relation nontrivial[24] (see Supplementary Section 1 for details).

Interestingly, the pump introduces a sub-micron-scale spatial inhomogeneity into the probe scattering response. Figure 1d shows a topographic image of a selected region of the heterostructure recorded by atomic force microscopy. The surface is clean and flat (nominal flake thickness: 110 nm) aside from folds at the borders of the investigated area. The simultaneously recorded broadband mid-IR near-field intensity scattered off the flake is small relative to the SiO$_2$ substrate and roughly uniform prior to near-IR photoexcitation (Fig. 1e, top). In contrast, near-IR pumping leads to a drastic increase in mid-IR scattering efficiency accompanied by a distinctive stripe pattern that runs parallel to the flake edges (Fig. 1e, bottom), a characteristic reminiscent of interference fringes due to surface plasmon and phonon polaritons on graphene[2,3,10,12] and hexagonal boron nitride[5,14], respectively. We also observe similar patterns on many other photoexcited BP heterostructures (see Supplementary Section 2).

In order to understand the microscopic origin of this photo-response we spectrally analyse the scattered mid-IR signal originating from the heterostructure using Fourier transform infrared spectroscopy (FTIR). Figure 2a shows the near-field FTIR amplitude spectra $s_3$ (for further details see Methods) measured over a flat gold reference surface (red), the bare $SiO_2$ substrate (blue), and the unexcited heterostructure (black). The scattering efficiency of gold in the mid-IR is spectrally flat, so the corresponding FTIR amplitude measured over gold represents a convolution of the input spectrum with the scattering response of the tip. The amplitude measured over $SiO_2$, on the other hand, is dominated by its longitudinal optical (LO) surface phonon polariton at 34 THz, as observed previously[4]. In contrast, multiple peaks are visible in the response of the unexcited $SiO_2$/BP/$SiO_2$ heterostructure (Fig. 2a, black curve). Here, each of the surface phonon-polariton modes occupying the Reststrahlen band of $SiO_2$ (31.9 THz to 37.4 THz) is energetically split into combinations of the interface modes at the top and bottom surfaces of the thin $SiO_2$ cover layer and the 300-nm-thick $SiO_2$ substrate layer. These mixed surface modes are shown in Fig. 2b (left, between horizontal white dashed lines), where we have calculated the dispersions of all the polaritons in the heterostructure using the transfer matrix formalism (see Supplementary Section 3 for details).

The simulations in Fig. 2b and Supplementary Section 3 further reveal that the finite thickness of the BP flake (110 nm) leads to an analogous splitting of its surface plasmon polariton (SPP): for low carrier densities (Fig. 2b, left), the surface plasmons at the two BP/$SiO_2$ interfaces combine to form distinct symmetric (lower energy, SPP$^-$) and antisymmetric (higher energy, SPP$^+$) branches. Sufficiently intense near-IR photoexcitation can shift the BP plasma frequency into the Reststrahlen band of $SiO_2$, leading to a coupling between the surface phonon and surface plasmon modes of the heterostructure (Fig. 2b, right). We can directly measure the corresponding spectral changes in the scattered near-field amplitude. To this end, the near-field tip is positioned on one of the bright stripes in Fig. 1e (indicated by a black circle) and the spectral amplitude $s_3$ and relative phase $\varphi_3$ of the scattered probe pulses are recorded as a function of delay after photoexcitation using time-resolved near-IR pump / mid-IR probe FTIR. Normalizing the complex scattering response at a particular pump-probe delay with the response at negative delay times reveals the photo-induced changes to the amplitude (Fig. 2c, left) and phase (Fig. 2c, right). Specifically, near-IR excitation enhances (suppresses) the amplitude of scattered radiation below (above) the unsplit phonon frequency (~34 THz). These changes manifest themselves in the phase spectra as a characteristic dispersive shape.

This response is consistent with the formation of a hybrid phonon-plasmon interface mode, as shown in Fig. 2b (right). When the BP plasma frequency enters the $SiO_2$ Reststrahlen band, the higher energy surface phonon branches are suppressed (see Supplementary Section 4). Meanwhile, the lowest lying LO surface phonon polariton (LO$^-$) couples to the upper branch of the split surface plasmon polariton (SPP$^+$), resulting in a large enhancement of the oscillator strength of the hybrid mode within the Reststrahlen band. According to our calculations, this amplification occurs close to the intersection point of the two bare surface polaritons, at a frequency that agrees well with the peak of the enhanced scattering response in Fig. 2c.

The propagation of these dominantly phonon-like polaritons explains the emergence of the stripe pattern in Fig. 1e. The sharp near-field tip provides the required momenta in the evanescent field at its apex[24-28], and thus launches mid-IR surface polaritons. After reflecting off the edges of the sample structure, they create standing waves with periodicity of $\lambda_{polariton}/2$, which can be resolved as a spatial modulation of the scattered near-field intensity. While such interference fringes are absent in near-field images of the unexcited BP heterostructure (e.g. Fig. 1e, top), they appear as soon as the near-IR pump pulse photo-activates the hybrid mode within the Reststrahlen band (e.g. Fig. 1e, bottom). The mode activation is not critically dependent on the near-IR pump photon energy provided the latter exceeds the BP bandgap (see Supplementary Section 1).

We test our assignment of the fringes (Fig. 1e) to the hybrid polariton mode by recording FTIR spectra as a function of the location of the near-field tip with respect to the standing wave pattern of the polariton – a technique that has been well established for mapping out the dispersion of surface polaritons[5,10].

Figure 3a depicts the near-field FTIR amplitude spectra of a photo-activated interface polariton on a typical flake measured along a line perpendicular to the fringes at $t_{pp}$ = 250 fs. The resulting hyperspectral plot clearly shows the periodicity of the interference fringes as a modulation in the FTIR amplitude. A one-dimensional Fourier transform along the tip scan direction converts Fig. 3a into an energy-momentum map of the hybrid interface polariton excited in our experiment (Fig. 3b). As expected, this mode is confined in energy to a region much smaller than our probe-pulse bandwidth (Fig. 2a) and in momentum to a region much smaller than the bandwidth supported by the evanescent field at the tip apex[4]. Moreover, the frequency and momentum of the polariton correspond well to the region of enhanced oscillator strength in the hybrid mode dispersion predicted by the transfer matrix model (Figs. 3b and 3c).

The hybrid phonon-plasmon-polariton mode is based on a photo-excited intrinsic semiconductor, and is thus both transient and completely switchable. Additionally, the fact that both the frequency and the wavevector of hybrid polaritons within the Reststrahlen band are only weakly dependent on the BP plasma frequency makes them particularly promising for ultrafast switchable nanooptics. To corroborate this key advantage, we record near-field snapshot images as a function of near-IR pump / mid-IR probe delay time (Fig. 4a). Although the mid-IR probe pulses are broadband, the tip launches relatively narrowband polaritons because their frequency and momentum are selected by the hybrid mode. Clear interference fringes emerge even within the rise time of the scattering response ($t_{pp}$ = 0 fs, defined in Fig. 1c). We investigate the emergence time of the fringes, and thus the onset time of the polariton mode, by resolving line scans of the scattered near-field intensity perpendicular to the edge of the BP flake in another, similar heterostructure as a function of pump-probe delay. As can be seen in Fig. 4b, the average amplitude of the polariton interference fringes in Fig. 4b rises from zero to above the noise level within < 50 fs (Fig. 4c). At later delay times a saturation of this amplitude suggests a rise from zero to half its maximum on the order of ~90 fs.

Generally, the fringes remain visible for the entire 5 ps lifetime of the enhanced scattering response (Fig. 4a) and, remarkably, the fringe spacing is nearly *constant* as a function of pump-probe delay despite the concurrent evolution of the BP carrier density and surface plasmon dispersion (see

Supplementary Section 5). The fringe spacing is similarly decoupled from the near-IR pump fluence (see Supplementary Section 4), indicating that the location of the enhanced oscillator strength within the Reststrahlen band is relatively insensitive to the carrier density, in contrast to the bare surface plasmon dispersion. Together, Figs. 2b, 2c, 3b and 4 illustrate an ultrafast switchable interface mode that occupies a confined region in energy/momentum space for the entire lifetime of the plasma. Our real-space data show good agreement with theoretical curves based on completely undamped polariton waves that only decay due to geometrical dispersion (see Supplementary Section 6). This long propagation length can be attributed to the phonon-like nature of the hybrid surface polariton within the Reststrahlen band[5,27]. Additionally, it has been shown previously that hybrid phonon-plasmon-polariton modes can exhibit propagation lengths surpassing those of their constituents[14].

In conclusion we have performed ultrafast measurements of BP on the nanoscale and resolved switchable interface polaritons in $SiO_2$/BP/$SiO_2$ heterostructures. These modes, which show constant energy and momentum and excellent coherence, present an exciting opportunity for optoelectronic devices. For example, a grating incorporated onto a BP heterostructure should allow for efficient incoupling to transient interface polaritons in the BP/$SiO_2$ heterostructure throughout their entire lifetime. Future devices could make use of the tunable bandgap of BP[15] to control its mode-activation energy as well as resonant near-IR excitation to achieve sub-optical-cycle switching times[29]. The in-plane asymmetry of BP[16] may also allow for further tunability, and in particular directional mode splitting. Looking forward, complex heterostructures that combine BP with graphene, transition metal dichalcogenides[30] and hexagonal boron nitride have the potential to provide a robust technological platform for polariton-based mid-infrared optoelectronics.

## Methods

### Sample preparation

The $SiO_2$/black phosphorus/$SiO_2$ heterostructures were produced by mechanical exfoliation of black phosphorus onto an intrinsic silicon wafer (380 µm) capped with a buffer layer of thermal $SiO_2$ (300 nm). The sample was subsequently sputtered with amorphous $SiO_2$ using an Argon plasma with a pressure of $2.5 \times 10^{-3}$ mBar and RF power of 70 W. The approximate growth rate is 1 Å/s, resulting in a nominal oxide thickness of 5 nm on the sample.

### Ultrafast s-SNOM setup

The ultrafast near-field scanning microscopy setup is based on an ultrastable Er:fibre laser system. Two branches are frequency-tailored in nonlinear optical fibres and their output is used for difference frequency generation of 60 fs long mid-infrared probe pulses centred at 36 THz (1200 cm$^{-1}$). A third arm serves as a pump beam operating at 1560 nm (6410 cm$^{-1}$). We use a nonlinear fibre to compress the pump pulses to 40 fs full width at half maximum (FWHM). Both the mid-infrared probe pulses and the near-infrared pump pulses are focused onto the tip of a commercial apertureless scanning near-field optical microscope using a parabolic mirror. The scattered mid-infrared probe intensity is measured using a mercury cadmium telluride detector. To suppress background radiation, we detect the scattered intensity $I_x$ demodulated at the x-th harmonic order of the tip tapping frequency. The corresponding amplitudes $s_x$ and phases $\varphi_x$ of the near-field spectra are obtained by Fourier transform infrared (FTIR) spectroscopy. For further details refer to ref. 24.

**Acknowledgements**

The authors thank M. Furthmeier for technical assistance, and A. Politano, S. I. Blanter, A. Chernikov, D. Peller, and M. Eisele for fruitful discussions. This work was supported by the European Research Council through ERC grant 305003 (QUANTUMsubCYCLE), and ERC grant 681379 (SPRINT), the Deutsche Forschungsgemeinschaft through Research Training Group GRK 1570, SFB 689 and Research grants CO1492/1 and HU1598/3.

**Author contributions**

M.A.H., M.S.V., T.L.C. and R.H. conceived the study. M.A.H., F.M., M.P., F.S., L.Z.K., T.L.C and R.H. carried out the experiment and analysed the data. L.V. and M.S.V. designed, fabricated and characterized the heterostructures of black phosphorus and silicon dioxide. M.A.H., F.M., M.P., T.F. and J.F. performed simulations. M.A.H., F.M., M.S.V., T.L.C and R.H. wrote the manuscript. All authors contributed to the discussions.

**Additional information**

Supplementary information is available in the online version of the paper. Reprints and permissions information is available online at www.nature.com/reprints. Correspondence and requests for materials should be addressed to M.S.V. or T.L.C.

**Competing financial interests**

The authors declare no competing financial interests.

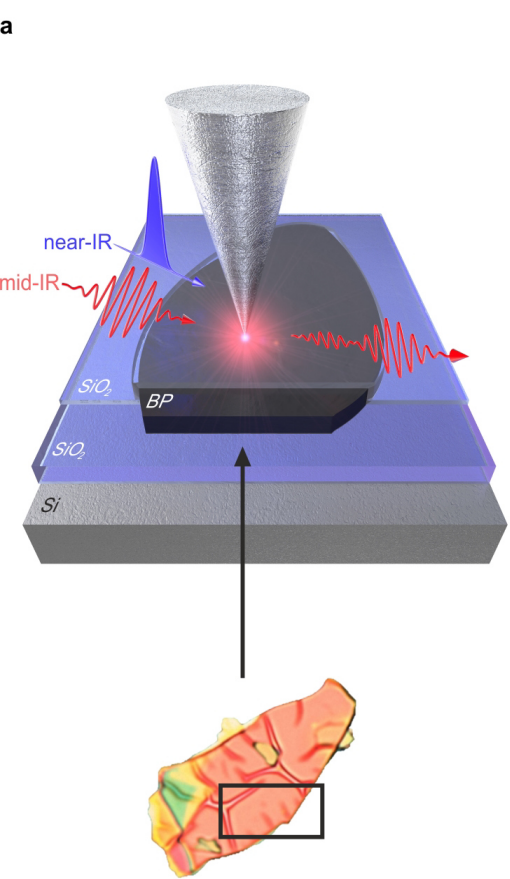
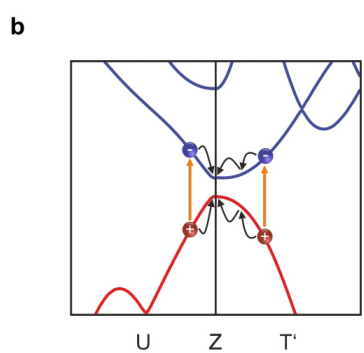
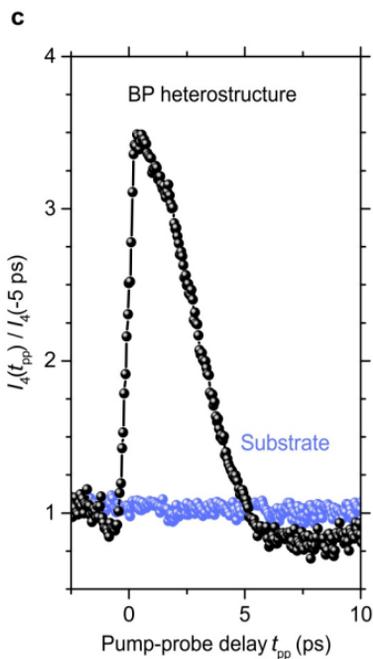
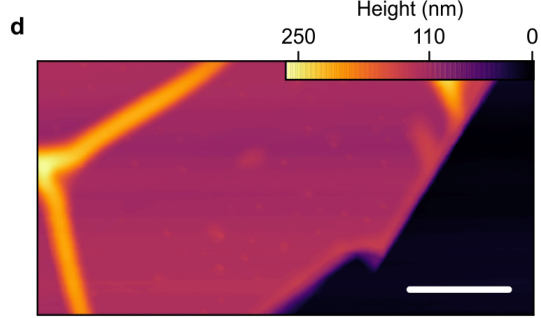
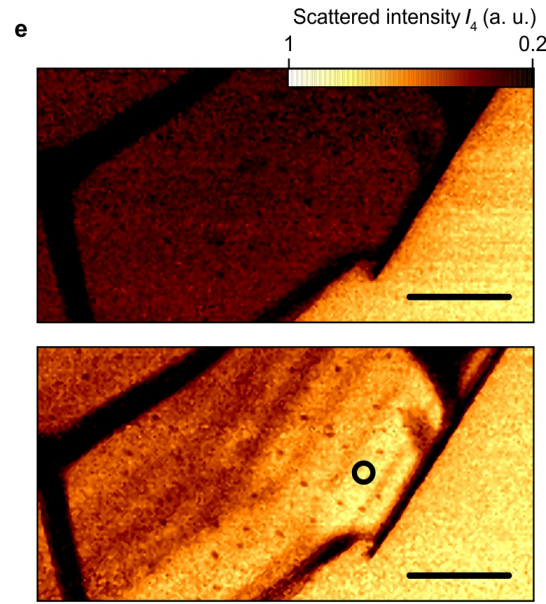

**Figure 1 | Overview of near-field microscopy on a switchable phonon-plasmon-polariton device.**
**a,** Schematic of the setup (top panel). An ultrafast near-infrared (near-IR) pump pulse is focused onto a heterostructure of $SiO_2$ and black phosphorus (BP) on a $SiO_2$ capped Si wafer. The sample is probed in the near field with a mid-infrared (mid-IR) pulse coupled to a sharp metallic tip. Bottom panel: Optical image of a single BP flake capped with $SiO_2$. **b,** Band structure of BP adapted from ref. 17 including orange arrows where a near-IR pulse centred at a wavelength of 1560 nm can excite electron-hole pairs. The curved black arrows indicate carrier cooling towards the band extrema. **c**, Ultrafast pump-probe dynamics of the scattered near-field intensity $I_4$ normalized to signals at negative delay time. The $SiO_2$ substrate (blue points) shows no dynamics, whereas the $SiO_2$/BP/$SiO_2$ heterostructure (black points) features a strong pump-probe signal. **d,** Atomic force microscopy image of the marked area in the bottom of **a** (nominal BP flake thickness: 110 nm). **e,** Spatial dependence of the scattered near-field intensity $I_4$ with and without pump excitation. Top: unpumped sample; Bottom: sample at $t_{pp}$ = 250 fs, after excitation by a near-IR pulse with a fluence of 1 $mJ/cm^2$. The black circle indicates the position where the data shown in Fig. 2c was taken. All scale bars are 2 μm long.

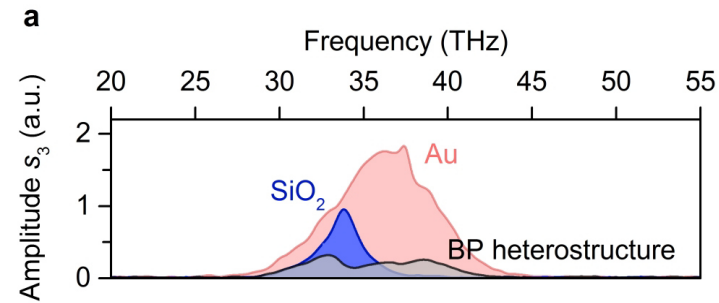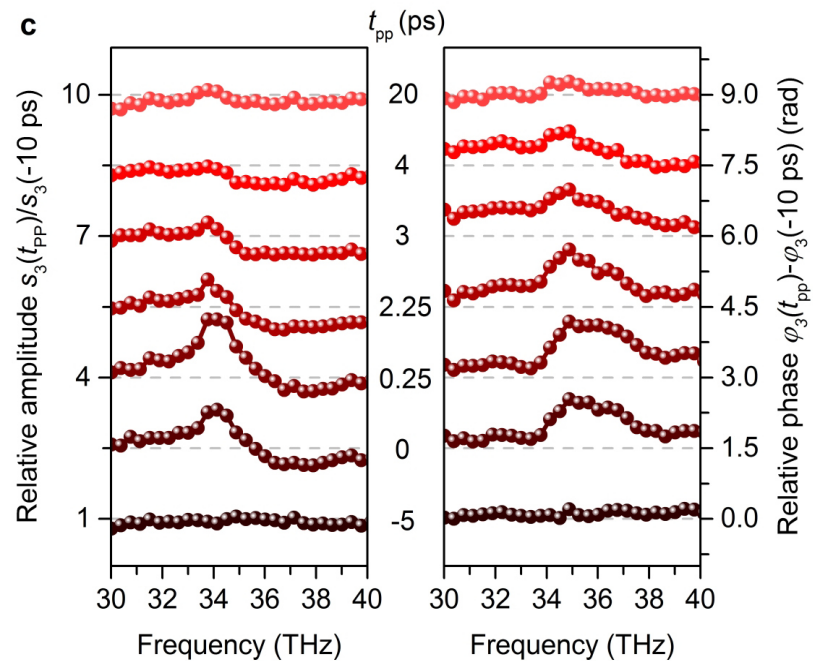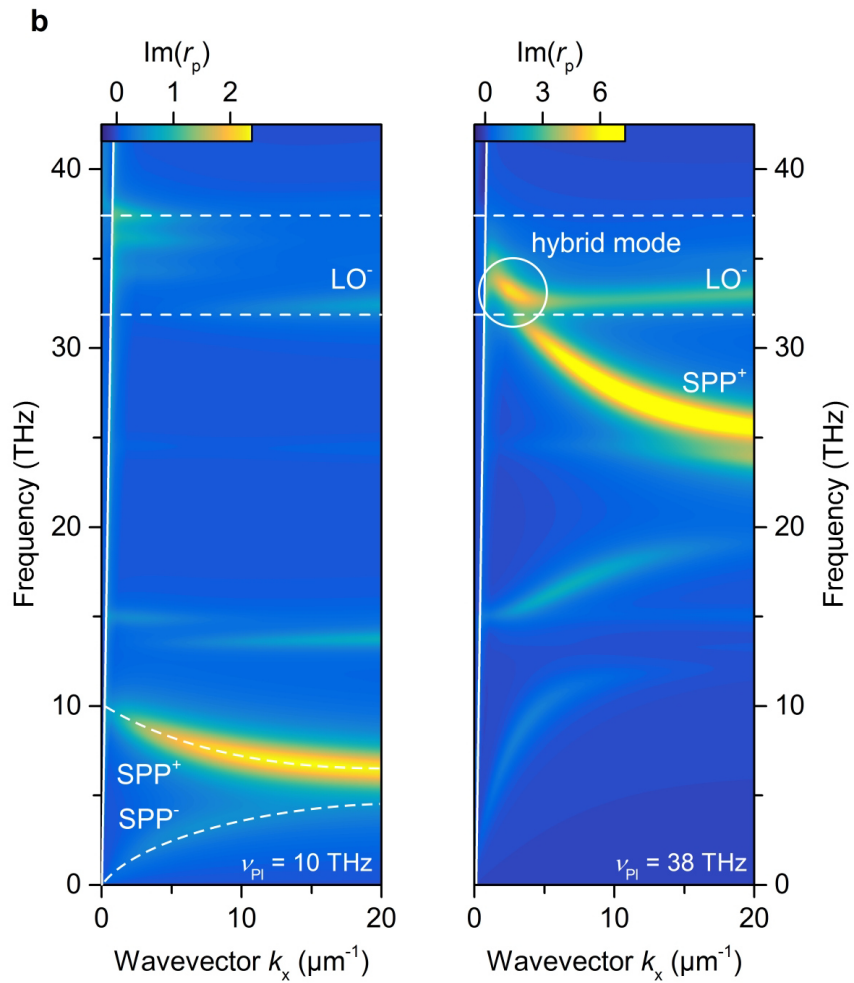

**Figure 2 | Ultrafast near-field spectroscopy of the SiO$_2$/BP/SiO$_2$ heterostructure. a,** Spectral amplitudes $s_3$ of the scattered near field from a gold reference sample, bare SiO$_2$ and the heterostructure from Fig. 1, resolved by FTIR. **b,** Theoretical dispersion relation of the polariton modes in the Air/SiO$_2$ (5 nm)/BP (110 nm)/SiO$_2$ (300 nm)/Si heterostructure, calculated by evaluating the imaginary part of the Fresnel reflection coefficient for p-polarized light $r_p$. At low BP carrier density (plasma frequency $\nu_{Pl}$ = 10 THz, left) the surface phonons in the Reststrahlen band of SiO$_2$ (the region between the horizontal white dashed lines) are split by coupling between surface phonon modes at different interfaces. We label the lowest energy phonon polariton in the band LO$^-$. At lower energy, the BP surface plasmon is split into two branches (SPP$^-$, SPP$^+$) by coupling between the surface plasmon modes at the two BP interfaces. At high BP carrier density ($\nu_{Pl}$ = 38 THz, right) coupling between SPP$^+$ and LO$^-$ produces a hybrid mode (white circle) within the Reststrahlen band. Solid white line: free-space light line. **c,** Ultrafast evolution of the spectral response of the SiO$_2$/BP/SiO$_2$ heterostructure. Amplitude (left) and phase (right) of the scattered near field are shown normalized to a spectrum at $t_{pp}$ = -10 ps. The sample was probed at the marked position in Fig. 1e.

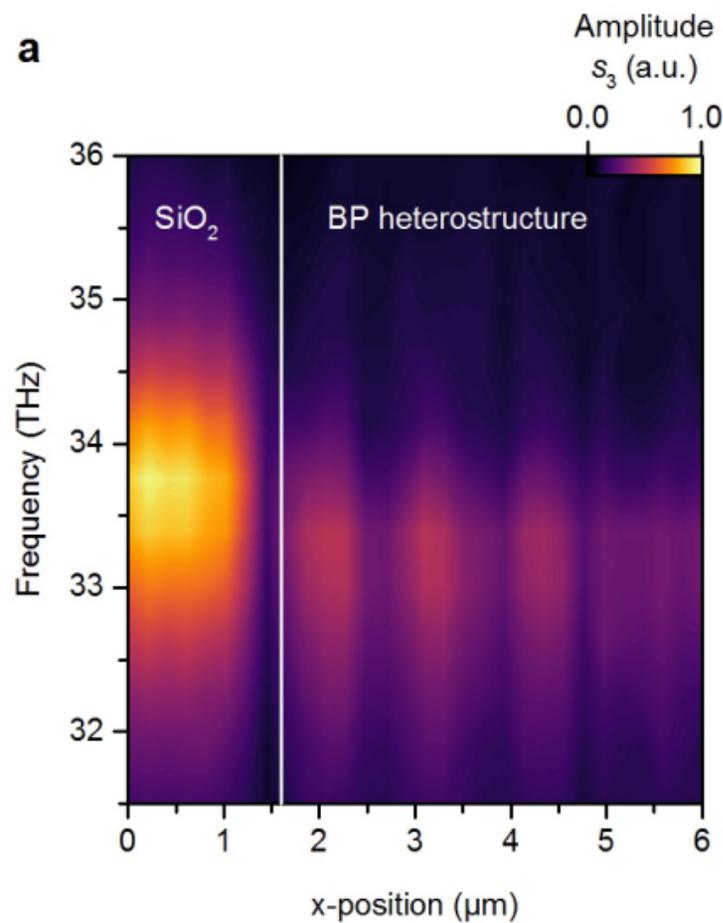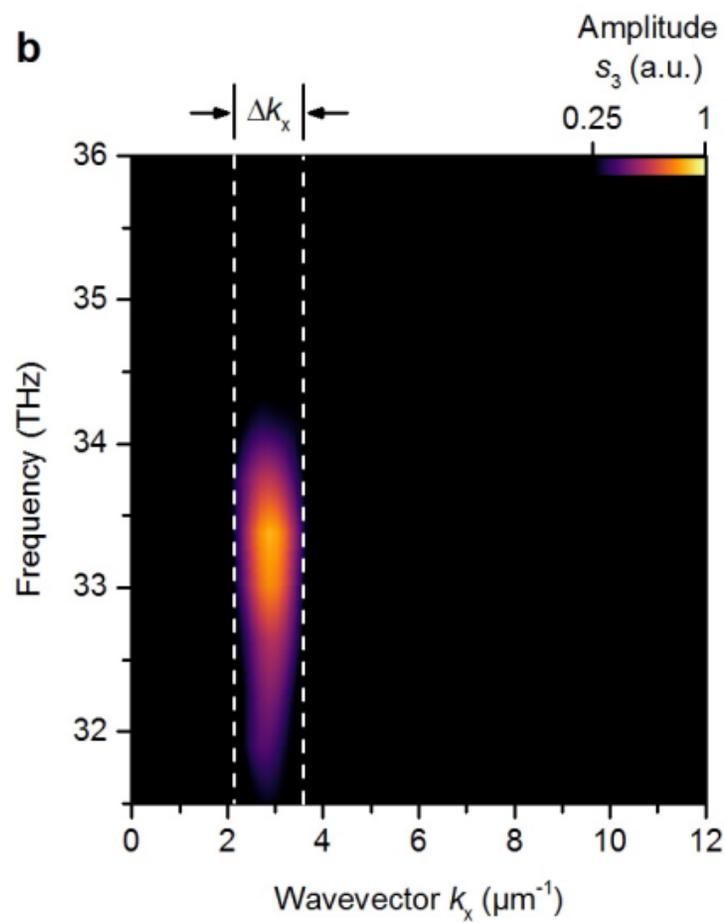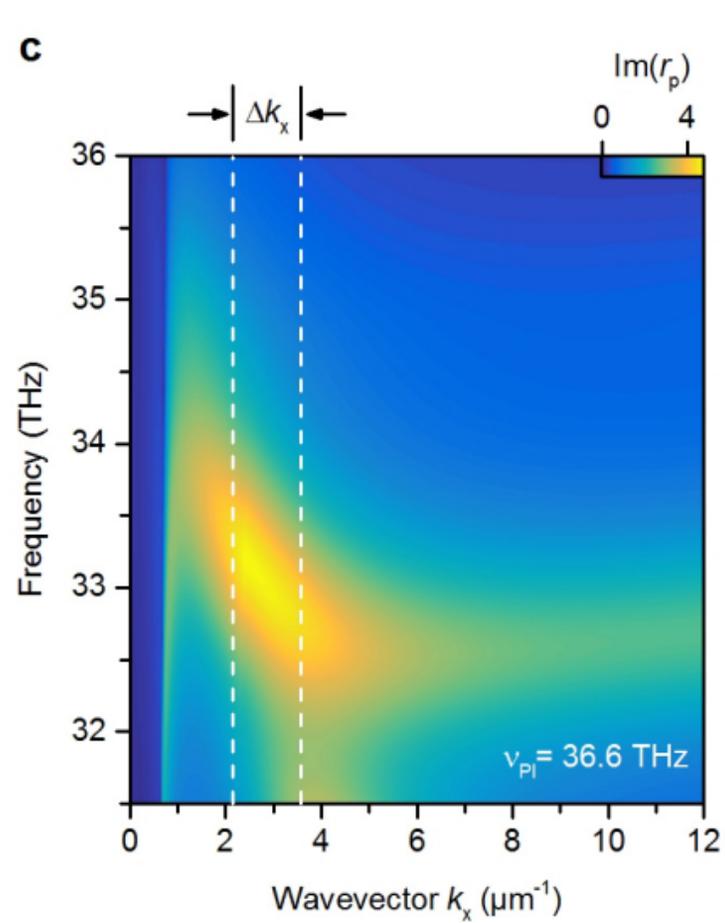

**Figure 3 | Hybrid phonon-plasmon polariton dispersion.** The dispersion of the hybrid mode in a SiO$_2$/BP/SiO$_2$ heterostructure can be obtained by a Fourier transformation of a hyperspectral map. **a,** Hyperspectral measurement, where the spectral amplitude $s_3$ of the scattered mid-IR probe is resolved by FTIR as function of tip position at a set near-IR pump / mid-IR probe delay of $t_{pp}$ = 250 fs. The tip is scanned from the SiO$_2$ substrate onto the heterostructure (left to right, BP flake thickness: 78 nm) perpendicular to the polariton interference fringes. The edge of the heterostructure is indicated by the solid vertical white line. **b,** A Fourier transformation of the hyperspectral data recorded on the heterostructure with respect to the tip position (x) yields the scattered amplitude $s_3$ as a function of the in-plane wavevector $k_x$ as well as frequency. We account for the experimental fact that the maxima in the standing wave pattern in **a** are separated by half the wavelength of the hybrid mode by dividing the wavevector axis by a factor of 2. A background subtraction analogous to the procedure described in ref. 10 was used before performing the Fourier transformation. The white dashed lines indicate the width of the minimum resolvable features in k-space, $\Delta k_x$, which is determined by the length of the hyperspectral map in real space. **c,** Theoretical dispersion of the hybrid mode, calculated via the imaginary part of the Fresnel reflection coefficient (as in Fig. 2b) for the Air/SiO$_2$ (5 nm)/BP (78 nm)/SiO$_2$ (300 nm)/Si structure. The white dashed lines in **c** bracket the same region of k-space as in **b**, highlighting the good agreement between the measured hybrid mode and the calculation.

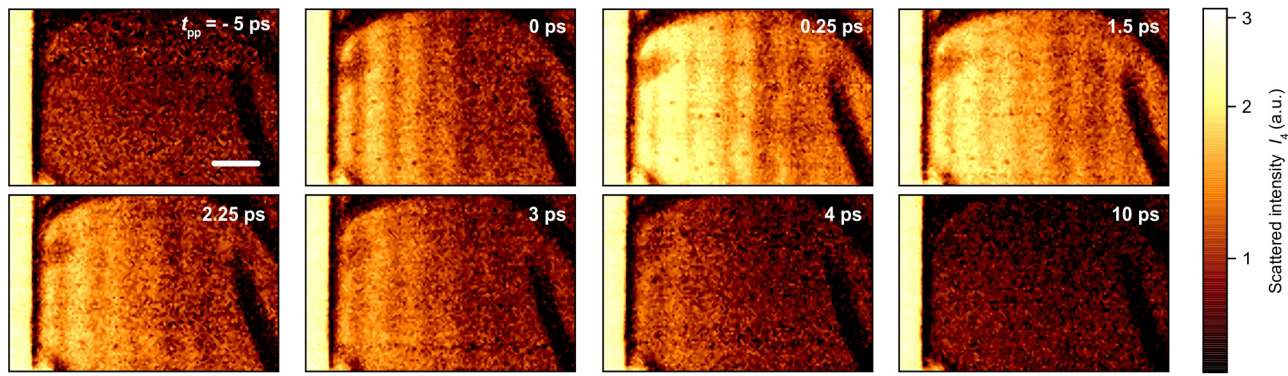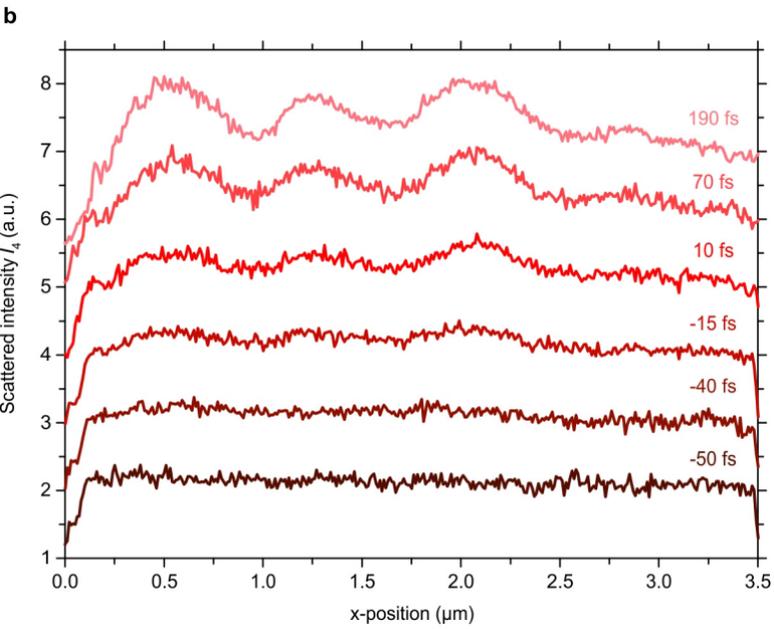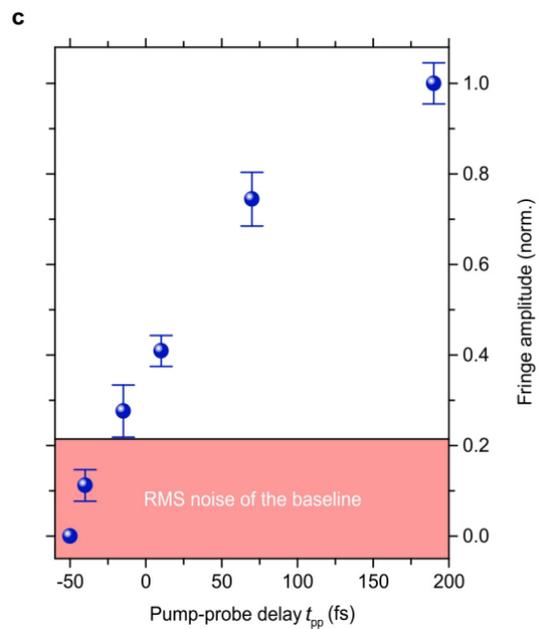

**Figure 4 | Switching and decay of a hybrid phonon-plasmon-polariton mode. a**, Scattered near-field intensity $I_4$ images of the heterostructure in Figs. 1 and 2, plotted for eight different delay times between the pump and probe pulses. The scale bar is 1 μm long and the incoupled mid-IR spectrum is shown in Fig. 2a. **b**, Onset of the polariton interference fringes on another heterostructure (thickness: 100 nm) following near-infrared photoexcitation, where the x-position is perpendicular to the BP flake edge. **c**, Average amplitude of the interference fringes in **b** as a function of pump-probe delay $t_{pp}$. Sine-curves were fitted to each individual fringe to extract their amplitudes, where the uncertainty of the fit routine is indicated by the error bars. The fringe pattern emerges within ~35 fs, as its amplitude surpasses the noise level (indicated by the red shaded region), which was determined by the root mean square (RMS) noise of the baseline for $t_{pp}$ = -50 fs. Subsequently, the fringe amplitude reaches half its estimated maximum value within ~90 fs.